\documentclass[%
 aip,
 sd,%
 amsmath,amssymb,
preprint,%
groupedaddress
]{revtex4-1}
\usepackage{pslatex}
\usepackage{mathbbol}
\usepackage{graphicx}
\usepackage{dcolumn}
\usepackage[
breaklinks,
colorlinks,
linkcolor=blue,
citecolor=blue,
urlcolor=blue]{hyperref}

\begin{document}


\title{A systematic study of the role of  dissipative environment in regulating entanglement and exciton delocalization in the Fenna–Matthews–Olson complex}
\author{Luis E. Herrera Rodr\'iguez}

\author{Alexei A. Kananenka}
\affiliation{%
Department of Physics and Astronomy, University of Delaware, Newark, Delaware 19716, United States
}%
\email{akanane@udel.edu.}

\date{\today}

\begin{abstract}

In this Article we perform a systematic study of the global entanglement and exciton coherence    
length dynamics 
coherence in natural light-harvesting system Fenna--Matthews--Olson (FMO) complex across various parameters of a dissipative environment from low to high temperatures, weak to strong system-environment coupling, and non-Markovian environments. A non-perturbative numerically exact hierarchical equations of motions method is employed to propagate the dynamics of the system. We found that entanglement is driven primarily by the strength of interaction between the system and environment, and it is modulated by the interplay between temperature and non-Markovianity. In contrast, coherence length is found to be insensitive to non-Markovianity. Furthermore, our results do not show the direct correlation between global entanglement and the efficiency of the excitation energy transfer in the FMO complex.

\end{abstract}

\maketitle

\section{\label{sec:level1} Introduction }

Interactions between quantum systems produce correlations with no classical counterpart, and systems with such quantum correlations are called entangled. Entanglement is one of the most intriguing concepts in quantum theory. It has been widely investigated in many research areas.~\cite{laflorencie2016quantum,erhard2020advances} Entanglement is at the core
of many technologically important applications such as quantum information and computing.\cite{horodecki2009quantum,cao2019quantum} However,  entanglement is very fragile and hard to control  due to the interactions with the environment resulting in decoherence.\cite{steane1998quantum} 


Systems at room temperature are not expected to show quantum signatures, because high temperature
is often associated with the classical behavior. 
A decade ago, two-dimensional electronic spectroscopy (2D ES) of light-harvesting complexes 
and conjugated polymers showed the presence of, what appeared like, quantum coherence effects at physiological conditions.\cite{engel2007evidence, scholes2017using, kim2021quantum, panitchayangkoon2010long,collini2009electronic,collini2009coherent} Specifically, these effects were observed in
2D ES spectra of the Fenna--Matthews--Olson (FMO) complex,\cite{fenna1975chlorophyll} a natural
light-harvesting complex mediating the energy 
transfer in the green sulfur bacterium \textit{Chlorobium tepidum}. The FMO is a homotrimer, consisting of eight bacteriochlorophyll-a (BChla) molecules (will be referred to as sites) per monomer embedded into a protein scaffold.\cite{schmidt2011eighth} BChla 1, 6, or 8 can be excited  from the chlorosome at the antenna when the sunlight is captured.  Energy is then highly efficiently funneled through the complex between the sites until it reaches site 3, at which the excitation is transferred to the reaction center,\cite{adolphs2006proteins, renger1998ultrafast, louwe1997toward, list2013toward} where the energy is used for chemical reactions needed to maintain bacterial metabolism.\cite{hu2002photosynthetic, olbrich2011atomistic} 
FMO and other light-harvesting complexes used years of evolution to tune their structure to achieve the high efficiency. 
 Experimentally observed quantum-like effects were rationalized in terms of coherences between the excitons, the collective excitations of pigments, which were also hypothesized to contribute to the high efficiency of the excitation energy transfer.\cite{engel2007evidence,tempelaar2014vibrational, collini2009coherent,duan2015origin} 
It is now believed that the quantum-like signatures in 2D ES spectra are 
due to the collective interplay of vibrational states and interexcition coherences cannot play a determining role in the energy transport due to their fast decay (sub–100-fs at room temperature).\cite{cao2020quantum,fassioli2014photosynthetic, li2022interplay,chenu2015coherence}

Although quantum effects may be short-lived and may not play an important role in mediating the excitation energy transfer in the FMO complex, they are nonetheless present and not well-understood.  They may be fundamentally important for understanding the energy transfer in artificial light-harvesting complexes where the environment  can be different from physiological  conditions.  
For example, the energy transfer between 
 a platinum phthalocyanine donor  and a zinc phthalocyanine acceptor was found to be 3 times more efficient compared with Förster resonance energy transfer (FRET) mechanism. This was rationalized by the wave-like transport related to interexciton coherence.\cite{kong2022wavelike}
Moreover, 2D ES  measurements at low temperature (20 K) showed interexciton coherences persists out to 200 fs in the FMO complex, where at low temperature, is possible to disentangle the interexciton coherence from long-lived vibrational coherence.\cite{duan2022quantum}
 
 Excition delocalization is a quantum property which provides a physical interpretation of the spatial extent of the exciton.
Coherence length as a measure of excition delocalization~\cite{smyth2012measures} was 
investigated by Moix \textit{et al.}\cite{Cao2012localization} in the FMO monomer, showing that the environment and static disorder can  enhance the coherence  length, and have a maximum value as a function of temperature. Dutta \textit{et al.}\cite{dutta2019delocalization} studied the coherence length as well, but in the FMO system reduced to three sites, finding a complete delocalization in the reduced system for shot-time values and a small but not null delocalization for long time values.


Delocalization implies a degree of entanglement between sites. Entanglement in the FMO complex has been investigated before.
Different bipartite entanglement measures, such as logarithmic negativity and multipartite entanglement across partitions such as tangle were developed to quantify the entanglement in excitonic systems.\cite{fassioli2010distribution,olaya2011characterizing,caruso02010power,caruso2009highly,smyth2012measures} Br\'{a}dler \textit{et al.}\cite{Blader2010discord} calculated the quantum discord as a measure of correlation in the FMO complex at cryogenic and physiological temperatures. They found the exact equivalence between the single-excitation quantum discord and single-excitation relative entropy of entanglement.

Sarovar \textit{et al.}\cite{sarovar2010quantum} developed a true global entanglement measure and used it to study entanglement in the FMO complex. They calculated multipartite entanglement in the FMO complex dynamics using site 1 and 6 as initial states. They showed that a non-zero bipartite entanglement between the distant sites, and the existence of the residual global entanglement, which is  enhanced at low temperature. Furthermore, they illustrated how multiparitite entanglement can exist in a highly decoherent  environment, and propose the existence of this long-live entanglement in larger light-harvesting complexes, such as LH1 and LH2 of the purple bacteria. They were studied later by Smyth \textit{et al.}\cite{smyth2012measures}

Thilagam \textit{et al.}\cite{thilagam2012multipartite} showed that multipartite entanglement can last between 0.5 and 1.0 ps in the non-Markovian regime and highlighted the importance of the bath correlation time and  the reorganization energy for sustaining  long time coherence.

Gonz\'alez-Soria \textit{et al.}\cite{gonzalez2020parametric} investigated the influence of the
environment on the population and coherence dynamics of the FMO complex using relative entropy. They elucidated
the role of temperature on the entanglement and coherences. 
By studying the coherence dynamics with a superposition of states chosen as the initial excitation they revealed a longer-lived entanglement compared to a single state being initially excited, and showed how large reorganization energies can boost the bipartite entanglement at low temperatures.\cite{gonzalez2023temperature} Additionally, Delgado \textit{et al.}\cite{delgado2023quantum} calculated Meyer--Wallach entanglement, normalized negativity measures, and maximum fidelity with respect to the closest separable state. They focused on BChla 3, 4, and 8. 
They showed that the state of the subsystem, representing only sites 3, 4, and 8, is close to a fully separable state at most times, with some cases of non-vanishing degree of entanglement.


Despite the aforementioned recent efforts, it is still not clear how protein environment 
modulates the entanglement in natural light-harvesting systems. Furthermore, it is not clear whether inter-site coherences are meaningfully described by an entanglement measure or delocalization measures as was discussed by Smyth  \textit{et al.} \cite{smyth2012measures} 
To date, no systematic study of the relation between these two quantum features in the light-harvesting complexes over a broad range of environmental parameters has been performed. In this Article we study the global entanglement and coherence length dynamics in the FMO complex across various regimes including important cases of non-Markovian dynamics, 
high and low temperature, as well as strong and weak system-bath coupling. 


Many computational methods have been used to study the dynamics of the FMO complex in order to understand the role of the protein environment in the excitation energy transfer and its  high efficiency.\cite{lightharvsreview2023, reviewMennucci, reviewUlrich,fassioli2014photosynthetic,jang2008theory, rebentrost2009environment,dutta2017environment,Fleming2005spectrocopy2D,scholak2011efficient,li2021charge,cao2009optimization, mohseni2014energy,fassioli2010distribution} In such simulations the FMO complex is described with an effective excitonic Hamiltonian where 
 only the  $Q_y$ transition is taken into account, which mainly involves the HOMO $\rightarrow$ LUMO excitation of BChla molecules, \cite{sirohiwal2020accurate} and the interaction between BChla sites is approximated by the dipole-dipole interaction. The environmental degrees of freedom (bath)
are modeled as a collection of quantum or classical harmonic oscillators.\cite{ishizaki2009theoretical,ishizaki2010quantum,cotton2016symmetrical} 
Theoretical methods are ranging from perturbative methods based on quantum master equations, such as Redfield theory
to non-perturbative approximate quantum-classical and numerically exact methods such as the Hierarchical equation of motion (HEOM) approach.\cite{tanimura2020numerically} Here we use the HEOM method because it properly
describes the non-Markovian effects that are known to be important in the dynamics of the FMO complex.\cite{ishizaki2009theoretical,chenu2015coherence,fassioli2014photosynthetic, tanimura2020numerically}

\section{Methods}
\label{section:methos}

\subsection{\label{sec:Model}Model}

In the single excitation manifold, energy transfer in the FMO system can be modeled by the common system-bath Hamiltonian    

\begin{eqnarray}
    \hat{H}= \hat{H}_s + \hat{H}_{sb} + \hat{H}_b,
\end{eqnarray}
where $\hat{H}_s$ is the exciton Hamiltonian, $\hat{H}_b$ describes the thermal reservoir or bath, and $\hat{H}_{sb}$ is the coupling  between the system and the bath. The exciton Hamiltonian takes the following form

\begin{equation}
    \hat{H}_{\mathrm{s}}=\sum_{n=1}^N \epsilon_n |n\rangle \langle n| +\sum_{k \neq n=1}^N J_{k n} |k\rangle \langle n|,
\end{equation}
where $\epsilon_n$ is the 
optical transition energy of the $n$th BChla site of the FMO complex with $n=\{1,\ldots,N\}$, and $J_{k n}$ are the excitonic 
couplings modeled by the dipole-dipole interaction between  $n$th  and  $k$th  BChla sites.

Dissipation and fluctuation in the system are induced by the interaction with the bath. The latter 
is modeled as a set of $N_b$ quantum harmonic oscillators
\begin{equation}
    \hat{H}_{\mathrm{b}}=\sum_{j=1}^{N_b}\left(\frac{\hat{p}_j^2}{2 m_j}+\frac{1}{2} m_j \omega_j^2, \hat{x}_j^2\right),
\end{equation}
where the momentum, position, mass, and frequency of the $j$th harmonic oscillator are given by $\hat{p}_j$ $\hat{x}_j$ $m_j$, and $\omega_j$ respectively. Each FMO site is assumed to couple to the same bath via 
\begin{equation}
    \hat{H}_{sb}=\sum_{n=1}^N \sum_{j=1}^{N_{\mathrm{b}}} c_j \hat{x}_j \hat{R}_n,
\end{equation}
where $\hat{R}_n= |n\rangle \langle n| $. The coupling constants $c_j$ are characterized by the spectral density
\begin{equation}
    J(\omega)=\frac{\pi}{2} \sum_{j=1}^{N_b} \frac{c_j^2}{m_j \omega_j} \delta\left(\omega-\omega_j\right).
\end{equation}

The reduced density operator for the system $\hat{\rho}_s(t)$  
is obtained by tracing out the bath degrees of freedom
\begin{equation}
    \hat{\rho_s}(t)=\operatorname{Tr}_{\mathrm{b}}[\hat{\rho}(t)],
\end{equation}
where $\hat{\rho}$ is the density operator for the total system.

When the bath follows Gaussian statistics, the  dynamics of the reduced density operator of the system
can be given in the form of the formally exact time-ordered influence functional. In interaction picture with respect to $\hat{H}_{\mathrm{s}} + \hat{H}_{\mathrm{b}}$ (indicated by the superscript $^{(I)}$) the reduced density matrix follows   
\begin{equation}
     \hat{\rho}^{(I)}_s(t) = \mathcal{T}_{\leftarrow} \exp \left\{ - \frac{1}{\hbar^2} \int_0^t dt_2 \int_0^{t_2} dt_1  \left\langle \mathcal{L}_{sb}^{(I)} (t_1) \mathcal{L}_{sb}^{(I)} (t_2)  \right\rangle _B \right\} \hat{\rho_s}(0),
\label{eq:funcional}
\end{equation}
where  $\mathcal{L}_{sb}^{(I)} \cdot = \left[ \hat{H}_{sb}^{(I)},\cdot \right] $ is a superoperator, $\left \langle \cdot \right \rangle_{B} = \operatorname{Tr}_{\mathrm{b}} \left[ \cdot e^{- \beta \hat{H}_b} \right] /Z $ is the quantum expectation value, $\beta = \left(k_B T\right)^{-1}$ is the inverse temperature,
($k_B$ is the Boltzmann constant), $Z = \operatorname{Tr}_{\mathrm{b}} \left[ e^{- \beta \hat{H}_b} \right]$ is the partition function, and $\mathcal{T}_{\leftarrow}$ orders $\mathcal{L}_{sb} (t)$'s in increasing time from right to left. 

The influence of the bath on the system is encoded in the second-order bath correlation function $C(\tau)$ which can
be expressed in terms of the spectral density as follows
\begin{equation}
    C(\tau) = \int_0^{\infty} d\omega\frac{J(\omega)}{\pi} \left[ \coth (\beta \omega /2) \cos (\omega \tau) - i \sin (\omega \tau) \right].
\end{equation}

Here the spectral density for the bath local to every BChla site is chosen to be an overdamped Drude--Lorentz spectral density\cite{weiss2012quantum}
\begin{equation}
    J(\omega) = 2 \lambda \frac{\gamma \omega}{\gamma^2+ \omega^2 },
\end{equation}
where $\lambda$ is the reorganization energy and $\gamma$ the cut-off frequency.

\subsection{\label{sec:Ent}Entanglement}


Entanglement between the sites in the FMO complex can be calculated from
its reduced density matrix. Usually, entanglement monotones are restricted to pure states or mixed bipartite states.\cite{wootters1998entanglement} Entanglement measures of multipartite mixed states are limited\cite{szalay2015multipartite} and hard to obtain analytically, given the difficulty of the convex roof construction,\cite{toth2015evaluating} which is an optimization problem over an infinite number of convex decompositions of the density matrix. Although the FMO complex is a multipartite system, it is often modeled with the
single excitation at a time. 
This enables a simple measurement to quantify the entanglement of a multipartite mixed system.  Following Sarovar \textit{et al.}\cite{sarovar2010quantum} we calculate 
the global entanglement for mixed states in the single excitation manifold as
\begin{equation}
    E[\rho] = - \sum_{n=1}^N \rho_{nn} \ln \rho_{nn} - S(\rho),
\label{equ:globlal_ent}    
\end{equation}
where $S(\rho)= - \operatorname{Tr} \rho \ln \rho $ is the von Neumann entropy and $\rho_{nn}$ is the corresponding diagonal element of the reduced density matrix. $E[\rho]$ measures all non-local quantum correlations between 
the excitonic states. Furthermore, because each chromophore in the FMO complex is treated as a two-level system, bipartite entanglement can be cast in terms of concurrences. The concurrence between 
sites $n$ and $k$ is given by

\begin{equation}
  C_{nk} = 2 |\rho_{nk}|.   
  \label{eq:concurrence}
\end{equation}

\subsection{\label{sec:Ex_c_l} Exciton Coherence Length}


Exciton coherence length, or the inverse participation ratio, is an important characteristic of excitons 
in molecular aggregates,  including photosynthetic systems.\cite{mukamel1997IPR,bourne2019structure} It measures the spatial extent of exciton delocalization.
The strong coupling between the sites in light-harvesting systems such as FMO leads to the 
delocalization of the exciton over many sites. In contrast, in the absence of system-bath interactions and negligible energetic disorder, the exciton can be delocalized over the whole complex. However, significant 
energetic disorder 
and the environmental influence lead to excition localization.  

Many definitions of the coherence length exist.\cite{dahlbom2001exciton,jiang2023unified} Here 
we employ the defintion introduced
by Meier \textit{et al.}\cite{meier1997multiple} which is based on the 
dispersion of the off-diagonal elements of the reduced density matrix in the excitonic basis  
\begin{equation}
    L_\rho=\frac{\left(\sum_{nk}^N\left|\rho_{nk}\right|\right)^2}{N \sum_{nk}^N\left|\rho_{nk}\right|^2} .
\end{equation}
Although this definition of the coherence length  tends to overestimate the delocalization compared to coherence measures, such as the tangle, it is still a very useful measure of the coherence length.\cite{smyth2012measures} To illustrate this definition we consider a few limiting cases. 
In the limit of the maximum incoherent mixing of pure states, for example at high-temperatures, 
$\rho= \mathbb{I} / N$,  coherences vanish, and the exciton is fully localized  giving $L_\rho=1$. In the
opposite limit of a fully coherent state $\rho= \mathbb{J} / N$, where $\mathbb{J}$ is the $N \times N$ matrix filled with ones, the exciton is fully delocalized over all sites giving  $L_\rho=N$. For a single pure state, such as the initial state assumed in the one-exciton manifold, all populations are zero except for the initial site yielding $L_\rho=1 / N$.

\subsection{\label{sec:Dataset}Data set}


To study the influence of the bath on the entanglement and coherence length in the FMO complex
we employ the previously generated by us database containing the FMO reduced density matrices calculated for a wide range of system and bath parameters.\cite{ullah2023qd3set} The data set was generated by solving Eq.~\ref{eq:funcional} using HEOM method.\cite{tanimura2020numerically}
PHI software was used for HEOM calculations.\cite{strumpfer2012open} The hierarchy truncation level and the number of Matsubara terms were systematically varied to
achieve the convergence of the populations to within 0.01 as detailed in Ref.\citenum{ullah2023qd3set}. 
The data set contains important cases of strong coupling and non-Markovian dynamics relevant to the realistic FMO system. 
The following combinations of bath parameters $\lambda= \{ 10,40,70, \cdots 520 \}$ cm$^{-1}$ $\gamma= \{ 25,50,75, \cdots 500 \}$ cm$^{-1}$ and $T= \{ 30,50,70, \cdots 510 \}$ K were used. A total of 879 entries (system-bath and bath parameters) are contained in the data set. Each entry is comprised of a set of time-evolved populations and coherences of the reduced density matrix up to 2 ps with the time-step of 0.1 fs. 
The data set was generated for the 8-site FMO Hamiltonian taken from Ref.\citenum{ke2016hierarchy} and also reproduced in Appendix (Eq.~\ref{eq:FMO_hamil}). 

\section{Results and Discussion}


Dissipation is controlled by the reorganization energy $\lambda$ which sets the system-bath coupling strength. For small reorganization energies the population and coherence dynamics are close to the unitary  dynamics.
The other bath parameter, cut-off frequency $\gamma$, sets the time scale of the bath dynamics and is associated with Markovian or non-Markovian nature of the environment.  Fast environments correspond to high cut-off frequency. In this case bath correlation functions decay exponentially and faster compared with the time scale of the system dynamics. Non-Markovianity is pronounced when the bath and systems' time scales are comparable, and the back-flow of information from the environment into the system is present. \cite{breuer2002theory,rivas2012open} Temperature is mainly associated with decoherence.

\subsection{Entanglement}
\label{section:entagle}


  We first investigate the global entanglement dynamics in the FMO complex for a fixed temperature, varying the reorganization energy and the bath cut-off frequency to characterize the interplay between strong system-bath interaction and non-Marokivanity in the entanglement dynamics. To enable a quantitative comparison, 
  we will calculate 
  the average energy gap in the FMO complex, defined as  $g=\frac{1}{N-1}\|H_s-\operatorname{Tr}(H_s) \mathbb{I} / N\|_*$, where $N=8$ corresponding to the eight sites of the FMO complex and  $\|X\|_*$ is the so-called nuclear or trace norm.\cite{srebro2005rank}  For the exciton Hamiltonian used here (Eq.~\eqref{eq:FMO_hamil}) $g= 157.17$ cm$^{-1}$. In what follows, for brevity,  the average energy $g$ will be referred to as energy. 
 
\subsubsection{\label{sec:global_ent} Global Entanglement dynamics}


\begin{figure*}
\includegraphics[width=0.99\textwidth]{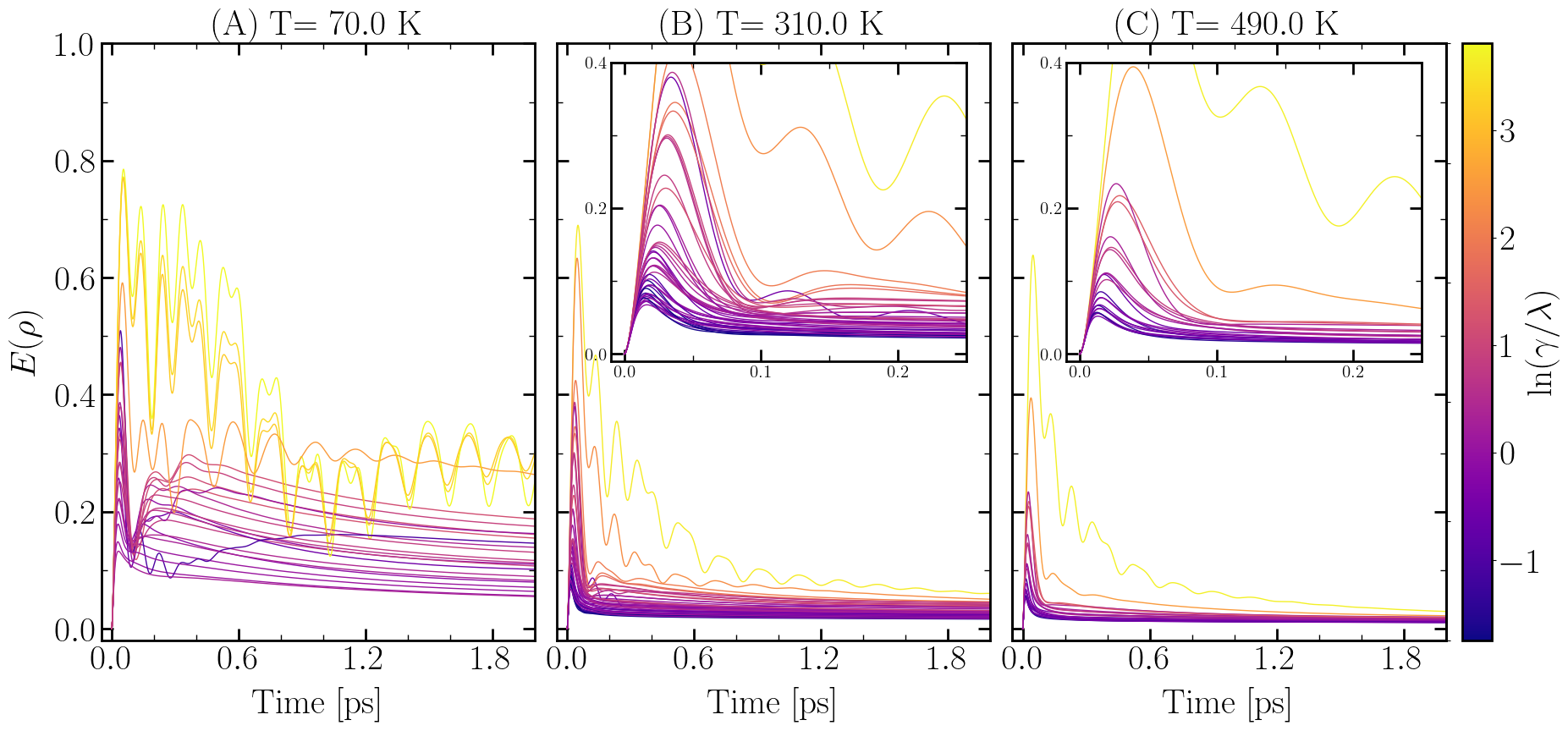}
\caption{\label{fig:glob_ent_temp}Global entanglement dynamics (Eq.~\ref{equ:globlal_ent}) of the FMO complex for different  reorganization energies and cut-off frequencies, for three temperatures (A) 70 K, (B) 310 K and (C) 490 K. Each line represents the global entanglement for the reduced density matrix, and the color of each line represents the dimensionless parameter $\ln (\gamma/ \lambda)$ which describes the interplay between dissipation and non-Markoviantiy. Entanglement in the high-temperature (B and C) regime is modulated by $\gamma/ \lambda$, while in the low temperature, two regimes of dynamics are found (A). Inset shows the early dynamics. Realistic  parameters of the natural FMO complex correspond to $ 0.35 < \ln (\gamma/\lambda) < 1.5 $.}
\end{figure*}

The global entanglement for various $\gamma/\lambda$ ratios, known as the Kubo number in the context of stochastic processes,\cite{goychuk2005quantum}  and fixed temperature, as a function of time is shown in Fig.~\ref{fig:glob_ent_temp}. In the high-temperature case, the parameter $\gamma/\lambda$ controls the overall shape of the global entanglement, as can be seen in Fig.~\ref{fig:glob_ent_temp}B,C. At $T=490$ K and $T=310$ K entanglement is highly suppressed by decoherence for small   $\gamma/\lambda$. In this case
the environment is highly non-Markovian and the exciton is strongly coupled to the bath which leads to 
localization. 

\begin{figure*}
\includegraphics[width=0.99\textwidth]{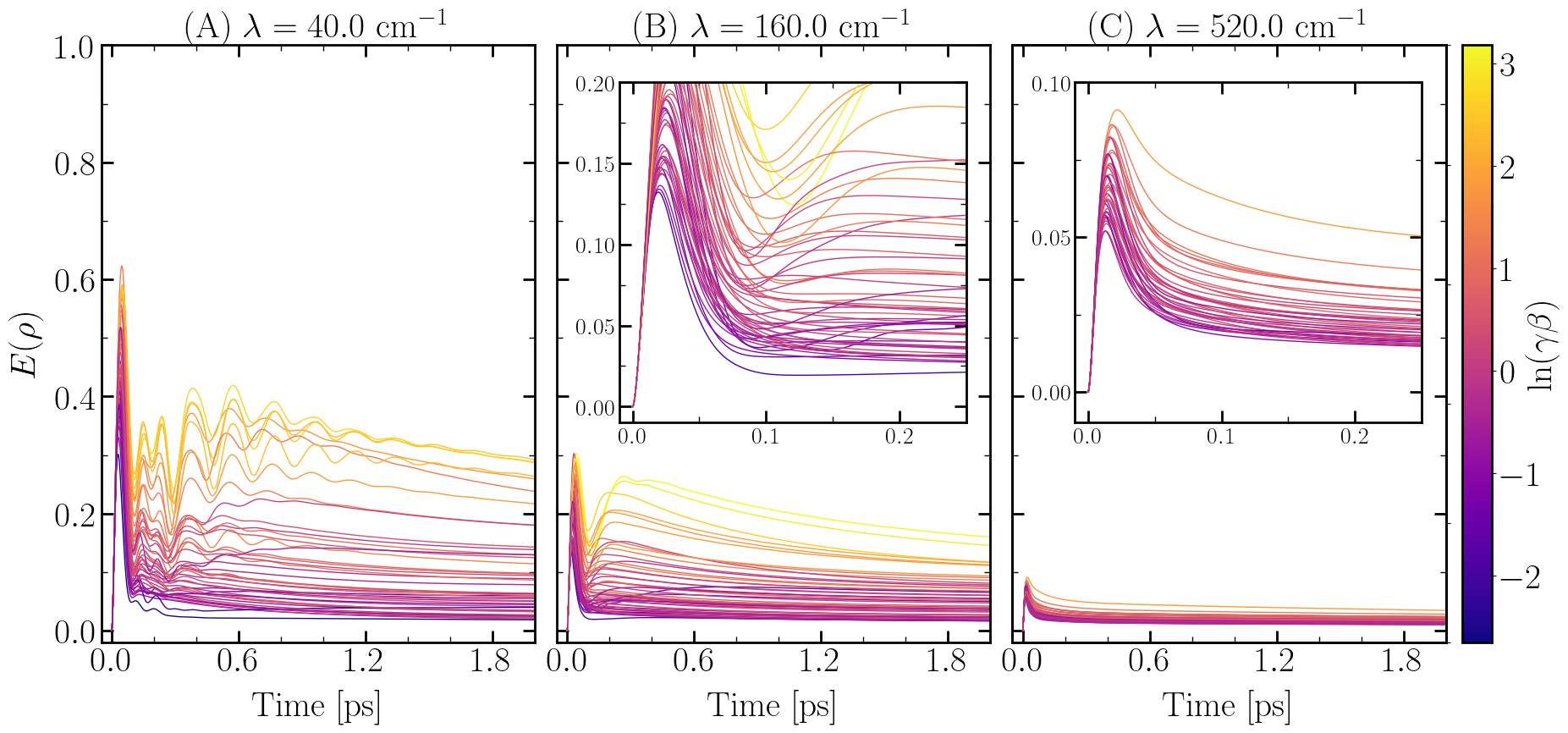}
\caption{\label{fig:global-lambda} Global entanglement dynamics (Eq.~\ref{equ:globlal_ent}) of the FMO complex for different temperatures and cut-off frequencies, for three reorganization energies (A) 40 cm$^{-1}$, (B) 160 cm$^{-1}$ and (B) 520 cm$^{-1}$. Each lines represents dimensionless parameter $\ln (\gamma \beta )$ pointing the interplay between decoherence and non-Markoviantiy. For fixed reorganization energy $ \gamma \beta$ determines the entanglement in all regimes; the overall shape clearly dictated by $\lambda$. Inset shows the early dynamics. Realistic parameters of the natural FMO complex correspond to $ -1.4 < \ln (\gamma \beta) < 0 $.}
\end{figure*}


Non-Markovian effects on the system's entanglement  
show revivals. For example, at $T=490 $ K and small  $\gamma/\lambda$ ratios, global entanglement is monotonically suppressed after one oscillation. Such decoherence obstructs any information flow from the environment to the system, as anticipated in the case of Markovian dynamics. In contrast, at $T=310$ K, $\gamma/\lambda \approx 1$, and high temperatures $k_B T / g > 1$ the non-Markovian effects are not completely suppressed and the global entanglement exhibits oscillations. Additionally, when thermal energy is comparable to the energy scale of the FMO complex ($k_bT / g \approx 1$), the non-Markovian signatures are not suppressed at all by decoherence. 


In the opposite scenario, when the $\gamma/\lambda$ ratio is large and temperature is high, 
the entanglement dynamics are highly oscillating. In this case, the environment is Markovian, 
but such a small interaction with the bath creates decoherence-free environment for the system leading to
the delocalization. Furthermore, we observe that at long times the entanglement grows with increasing $\gamma/\lambda$.


In the low-temperature limit, the dynamics is much richer and more sensitive to the $\gamma/\lambda$ ratio. In such cases, decoherence effects are weaker, and the global entanglement oscillates notably for up to 2 ps for small reorganization energies. This can be understood by taking into account that the system at 2 ps may not be in equilibrium with the environment yet. Coherent oscillations of the populations are present in the FMO complex, indicating a significant delocalization of the exciton and fast transfer between the sites. When the ratio   $\gamma/\lambda$ is small and  temperature is low, the non-Markovian effects become notable in the entanglement dynamics. For some cases, entanglement shows an increase after a few oscillations, as expected from non-Markovian dynamics. However, after oscillating at early times it typically  monotonically decreases.



Moreover,  
in all cases the entanglement initially rapidly increases. At the early stages of the dynamics the exciton is mainly delocalized over sites 1 and 2 due to 
the relatively strong
coupling between these sites. As time progresses the  
other sites become populated, and due to the incoherent transport, entanglement decreases. Nevertheless, as we clearly show the entanglement could still persist or possibly even intensify. 



To summarize, at high temperatures, $\gamma/\lambda$ controls global entanglement and its residual value. In contrast, in the low-temperature regime, small decoherence enables stronger entanglement, 
which oscillates 
and is only weakly damped.  



We now discuss the interplay between temperature and cut-off frequency and how it affects the entanglement.  Fig.~\ref{fig:global-lambda} shows global entanglement for various  reorganization energies. We explore a weak coupling case when $\lambda = 40 $ cm$^{-1}$ which is close to experimentally relevant values  for the FMO complex,\cite{cho2005exciton,adolphs2006proteins,plenio2008dephasing} an
intermediate coupling $\lambda / g \approx 1$ and strong system-bath coupling. Additionally, in Appendix \ref{apendix:lowlamda}, we show the robustness of the global entanglement at low reorganization energies.



 
The progression of the entanglement dynamics is predominantly influenced by the parameter $\lambda$ across the entire spectrum of $\gamma \beta$, although there are minor variations for elevated values of $\ln(\gamma \beta) \gtrapprox 2$. The entanglement values are chiefly determined by the parameter $\gamma \beta$ for fixed reorganization energy. When dissipation is weak or moderate and temperatures are low, the
decoherence effects are not strong, and the correlation between sites persists over time, causing global entanglement to oscillate. In contrast, when $\gamma \beta$ is small, high temperature causes decoherence in the FMO complex, but the non-Markovian effects could lead to oscillations in the entanglement. In this regime  we associate these oscillations with the back-flow of information of the non-Markovian environment, despite moderate dissipation and relatively strong decoherence effects.


Examining the dependence of entanglement on the reorganization energy, we observe that the
strong dissipation suppresses the
entanglement after the latter rapidly increases at early times. Furthermore, the maximum value of entanglement decreases with stronger dissipation, as expected, given that the correlations among sites are lost into the environment. When the reorganization energy is comparable with the average energy $\lambda / g \approx 1$, 
oscillations persist longer.
The first oscillation is associated with fast exciton delocalization over sites 1 and 2 followed up by localization while the second oscillation could be linked to the delocalization of the exciton in other sites of the FMO complex and a slow localization over time.
In this case, because the exciton-bath interaction is comparable with the exciton energy, correlations between the sites are destroyed asymptotically after the second oscillation. Incoherent transport via bath may increase the correlation between sites.



For a realistic reorganization energy of the FMO complex, $\lambda = 40 $ cm$^{-1}$, the entanglement exhibits several oscillations which are
damped by decoherence at high temperatures. Nevertheless, as mentioned before, non-Markovian effects may still be present for some parameters of the environment, increasing the entanglement after some oscillations. 

\subsubsection{\label{sec:Bipar_ent} Bipartite Entanglement dynamics}

\begin{figure*}
\includegraphics[width=\textwidth]{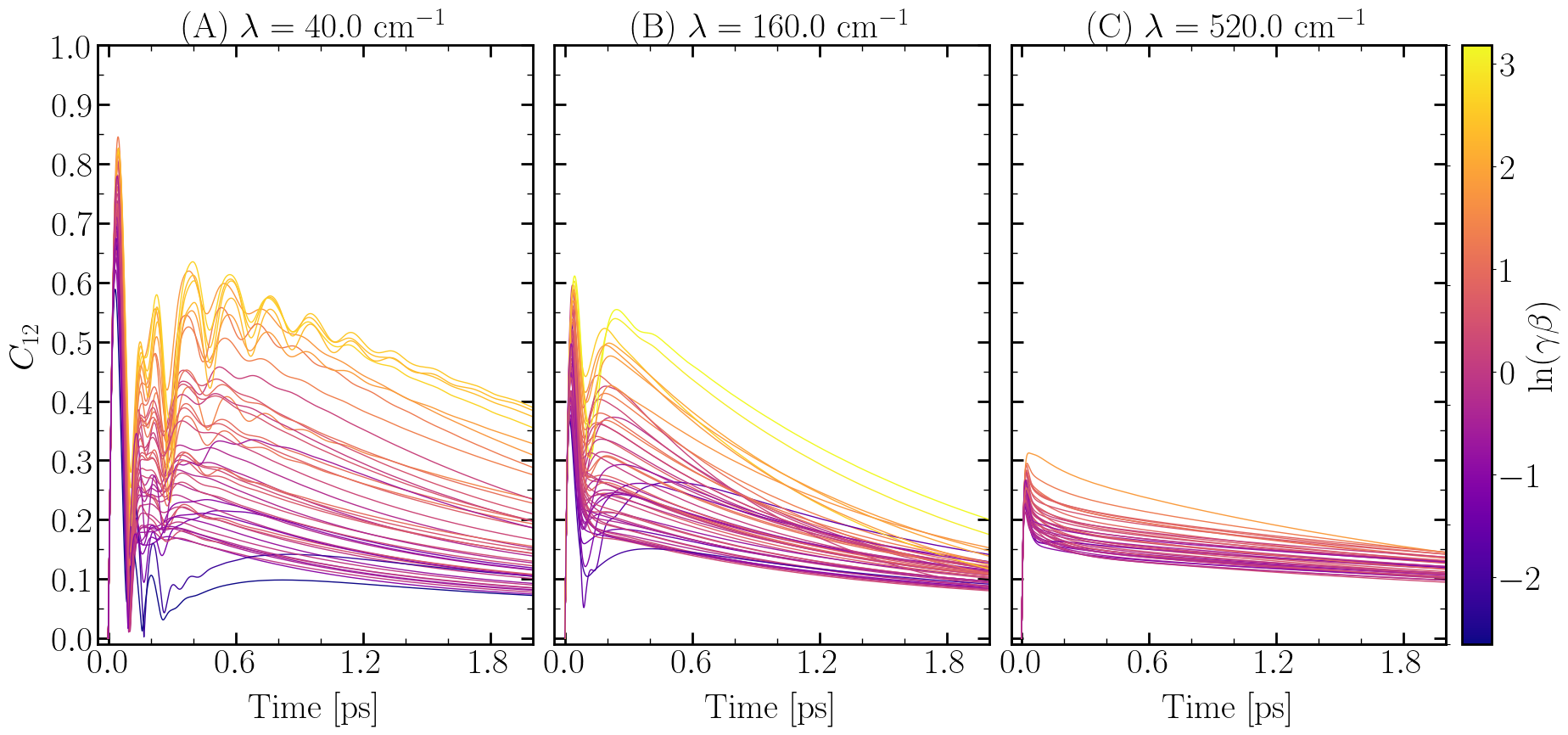}
\caption{\label{fig:bi_12}Bipartite entanglement dynamics between sites 1 and 2, $C_{12}(t)$,  in the FMO complex for different  values of the reorganization energy: (A) 40 cm$^{-1}$, (B) 160 cm$^{-1}$, and (C) 520 cm$^{-1}$. }
\end{figure*}


Global entanglement quantifies the quantum correlations among all sites in the FMO complex. Close examination of the correlations between pairs of sites
 reveals a non-trivial behavior of the bipartite entanglement. We now focus on the bipartite entanglement, measured with the concurrence 
 calculated using Eq.~\ref{eq:concurrence}, between sites 1-2, 1-3, and 3-4.  Sites 1 and 2 are the most strongly coupled sites $(|J_{12}| \approx 80 $ cm$^{-1})$, and site 1 is where the initial excitation occurs. Site 3 is known to be responsible for transferring the excitation to the reaction center, where the charge separation occurs. It is directly linked with the efficiency measures of the energy transport of the FMO complex. \cite{mohseni2014energy} Additionally, despite sites 1 and 3 being far apart  ($ \sim 28$ \AA), and weakly coupled $(|J_{13}| = 3.5 $ cm$^{-1})$, a non-trivial entanglement exists even among these sites.~\cite{sarovar2010quantum}


We first consider the concurrence between sites 1 and 2 shown in Fig.~\ref{fig:bi_12}, for fixed reorganization energies 
and scanning through $\gamma \beta$ values. We choose this parameter and not $ \gamma/\lambda $ because it better controls the maximum value of entanglement at short times. Different values of $\lambda$ cause different numbers of oscillations.


At the early stages of the dynamics the concurrence between sites 1 and 2 is the major contributor to the entanglement of the system. Its behavior is qualitatively similar to the global entanglement. It continues to oscillate, beyond
the first oscillation, for small reorganization energies, and it persistently oscillates when the 
reorganization energy is comparable to $g$. For strong dissipation the concurrence between sites 1 and 2  decays to the equilibrium value.

\begin{figure*}
\includegraphics[width=\textwidth]{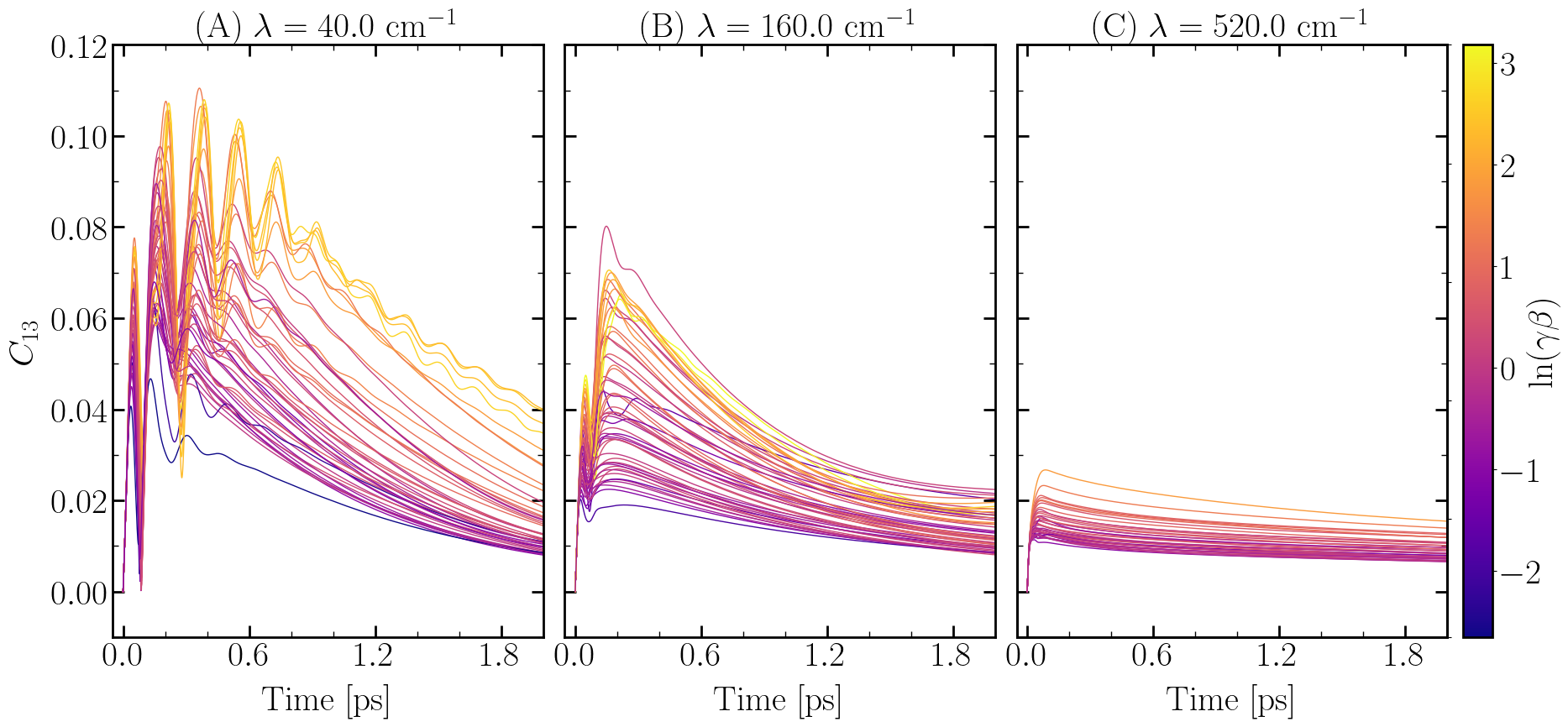}
\caption{Bipartite entanglement dynamics between sites 1 and 3, $C_{13}(t)$, in the FMO complex for different values of the reorganization energy: (A) 40 cm$^{-1}$, (B) 160 cm$^{-1}$ and (C) 520 cm$^{-1}$. }
\label{fig:bi_13}
\end{figure*}


Fig.~\ref{fig:bi_13} shows that bipartite entanglement between sites 1 and 3 is non-zero, as expected. In the  Fig.~\ref{fig:bi_13}B we show a remarkable feature of the concurrence between sites 1 and 3.  When $\ln (\beta \gamma)  \approx 0 \; (\beta \gamma \approx 1)$, the concurence may take higher values compared to cases when $\ln (\beta \gamma) $ is higher. 
It is tempting to related it to the high efficiency of the energy transfer. 
The excitation energy transfer efficiency can be qualitatively estimated with the help of the following parameter  $\Lambda = \frac{\lambda}{\beta \gamma g}$.\cite{mohseni2014energy} 
 $\Lambda$ close to 1 corresponds the optimal energy transfer efficiency.
 For $\lambda /g \approx 1$ and $\beta \gamma \approx 1$, as
in this case, $\Lambda \approx 1$ so the efficiency is close to 100 \%.  However, outside of this regime, most generally, the bipartite entanglement dynamics do not follow this behavior.   

\begin{figure*}
\includegraphics[width=\textwidth]{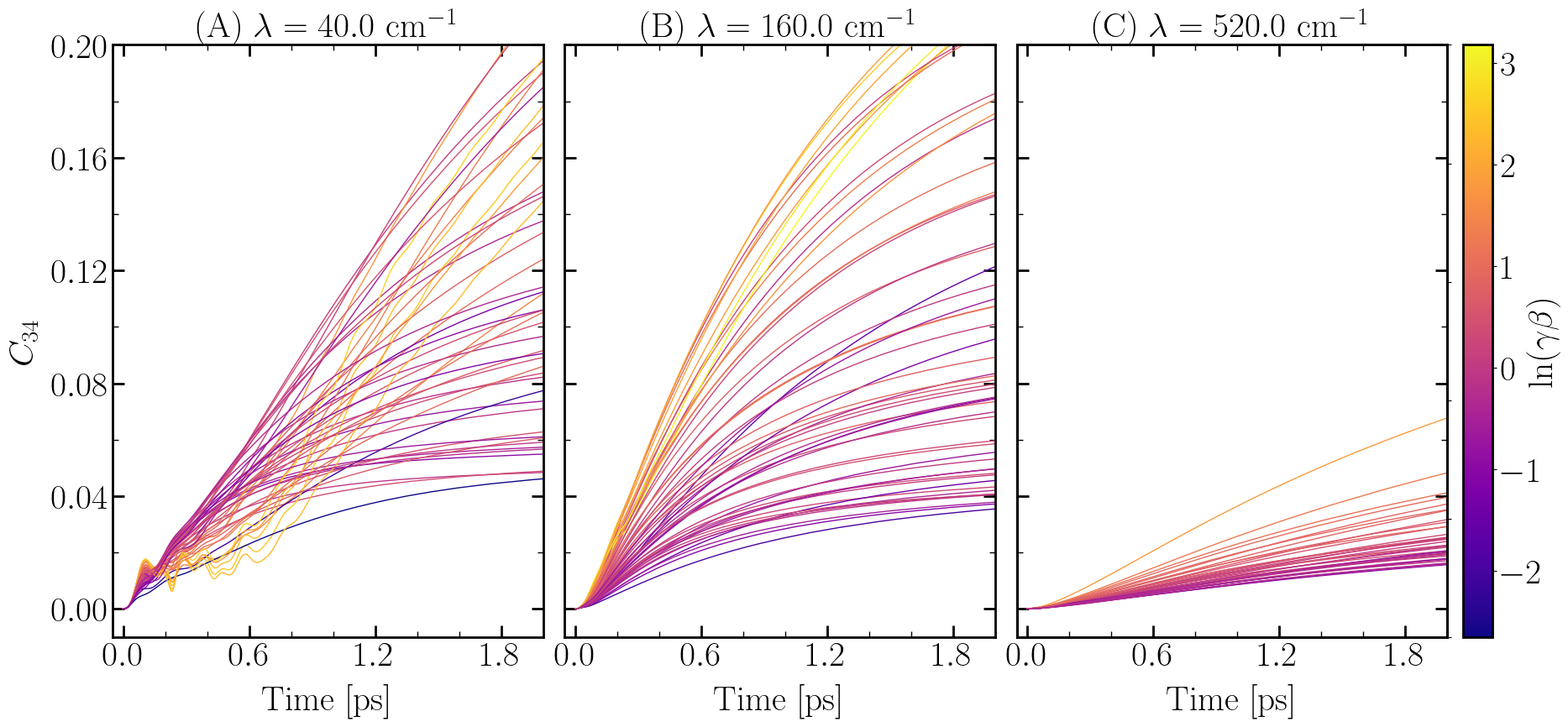}
\caption{\label{fig:bi_34}Bipartite entanglement dynamics between sites 3 and 4, $C_{34}(t)$, in the FMO complex for different values of the reorganization energy: (A) 40 cm$^{-1}$, (B) 160 cm$^{-1}$ and (B) 520 cm$^{-1}$.}
\end{figure*}


Finally, we examine the concurrence between sites 3 and 4 shown in Figure \ref{fig:bi_34}. $3\to4$ is the final step in the excitation energy transfer pathway.\cite{moix2011efficient} Bipartite entanglement increases as the excitation energy transfer process takes place. When the reorganization energy is high $\lambda > g$, the concurrence between sites is suppressed, which is 
expected from the complete decoherence and loss of entanglement. In contrast, bipartite entanglement is enhanced for a moderate to weak system bath interactions. In this case, the entanglement for $\lambda = 40$ cm $^{-1}$ and  $\lambda = 160$ cm $^{-1}$ is similar, which is in stark contrast to the concurrence between sites 1-2 and 1-3. This behavior is also observed in the bipartite entanglement between sites closest to site 3: pairs 3-4, 3-7, and 2-3. This may be related to the robustness of the excitation energy transfer
in the FMO complex: multiple pathways are available for the excitation to reach the target site. Again, when the reorganization energy is small, bipartite entanglement exhibits oscillatory behavior for high $\beta \gamma$.

\subsection{Coherence length}
\label{section:coherencelenght}

\begin{figure*}
\includegraphics[width=\textwidth]{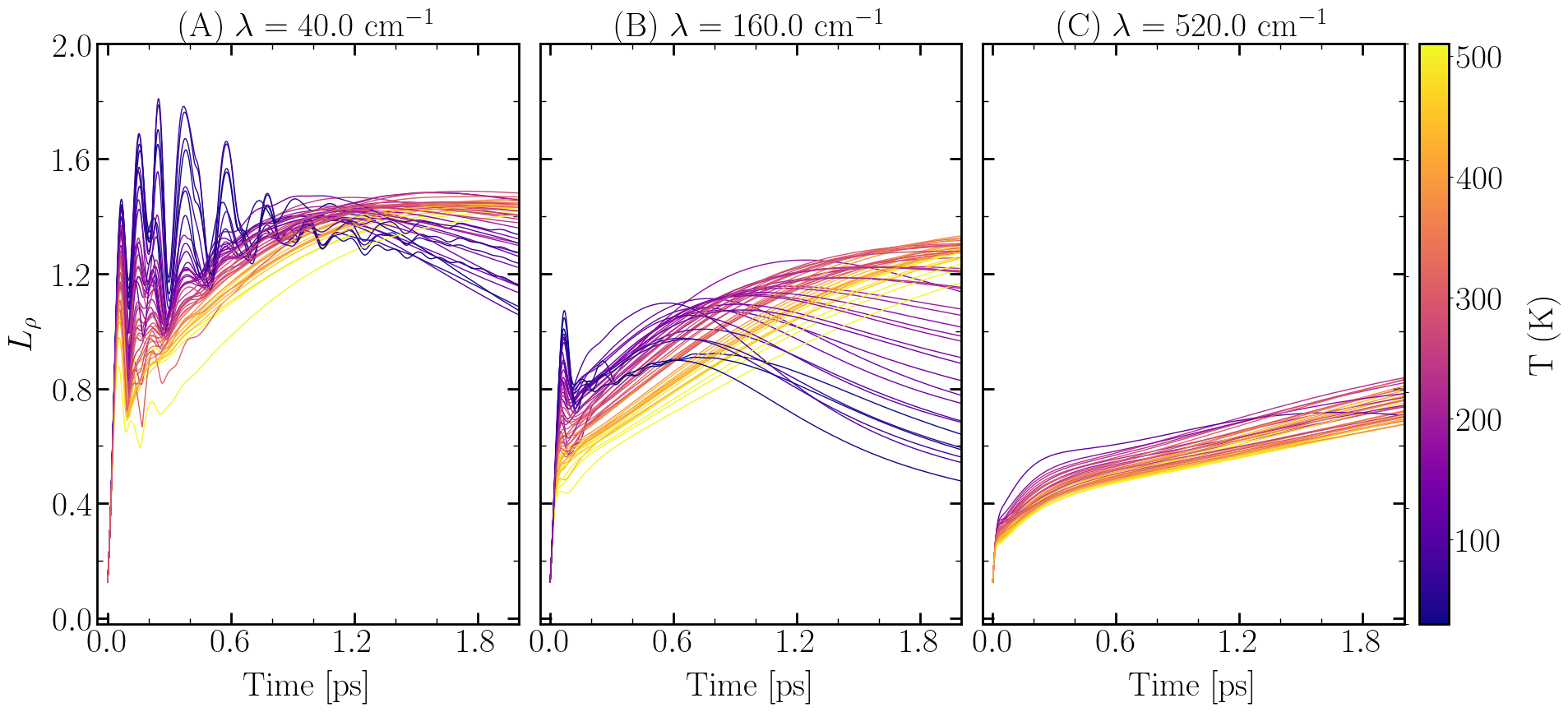}
\caption{\label{fig:corr_leng} Coherence length dynamics $L_{\rho}$ for fixed values of the reorganization energy $\lambda$  (A) 40 cm$^{-1}$, (B) 160 cm$^{-1}$, and (C) 520 cm$^{-1}$.}
\end{figure*}

We turn our attention to environmental impact on the coherence length. In contrast to entanglement, the coherence length is driven principally by the temperature, for a fixed value of reorganization energy, as illustrated in Fig.~\ref{fig:corr_leng}. The cut-off frequency controls the non-Markovian effects which are shown at early stages of the dynamics, these effects  seem not to be  relevant in the longer-time coherence length dynamics. In Fig.~\ref{fig:gamma_fix}\ref{apendix:cut-oof_ind}, we show an equivalent plot to Fig.~\ref{fig:corr_leng} but for a fixed cut-off frequency, showing qualitatively the same behavior, thereby suggesting that the cut-off frequency has no effect in the coherence length for the range of parameters chosen.

The coherence length phenomena in the FMO complex is similar to the one of a biased two-level system. In this scenario, the energy difference between the sites drives the excitation from the initial site to a target site.
In the absence of the vibronic environment, the excitation in the FMO complex tends to populate the lowest energy site. The exciton will not be significantly delocalized over all of the eight sites in principle, due to intrinsic energetic disorder.  As shown in Fig.~\ref{fig:corr_leng}, the coherence length for a high reorganization energy increases, for most values monotonically, potentially reaching 1. The coherence length between $1/N$ and 1 represents mixed states with different populations (i.e., $\rho \neq \mathbf{I}/N $), or mixed states over subspaces (some population(s) might be zero) and some subspaces might be  disconnected.

\begin{figure*}
\includegraphics[width=\textwidth]{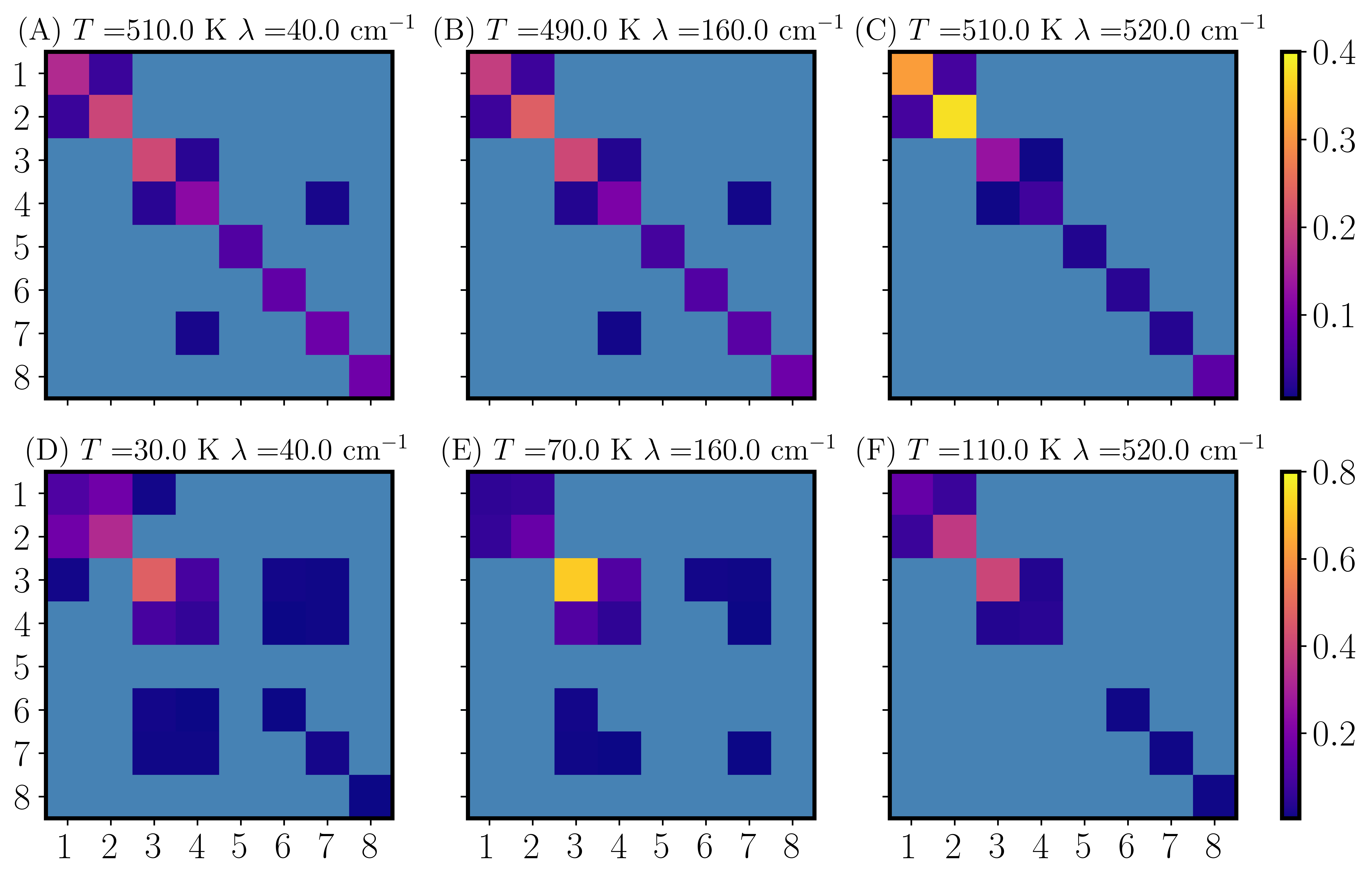}
\caption{\label{fig:dm} The absolute values of the elements density matrix of the FMO complex at 2 ps for different temperatures and reorganization energies of the bath. Colorbar represents the norm of the density matrix element $\rho_{nk}$, and values below $|\rho_{nk}| <0.005$ are set in the lightest blue color. Panels A, B, and C represent the high-temperature limit, and panels D, E, and F correspond to the low-temperature limit.  Note the difference in scales of the color bar for each limit. Panels A and D correspond to small reorganization energy, panels B and E to the moderate values, and panels C and F to large reorganization energy.}
\end{figure*}

The absolute value of all elements of the reduced density matrix at 2 ps are schematically shown in Fig.~\ref{fig:dm}. Fig.~\ref{fig:dm}C,F are for  the higher reorganization energy, and Fig.~\ref{fig:dm}C and Fig.~\ref{fig:dm}F represent the high and low temperature cases, respectively. Although they have roughly the same coherence length, they represent different situations. Due to the high temperature, the density matrix in Fig.~\ref{fig:dm}C will represent the classical mixture with zero coherences, and populations being   distributed over the sites, with sites 1 and 2 being mostly populated. In contrast, in Fig.~\ref{fig:dm}F, a separation between some sites is shown, but coherence is presented in one subspace. Populations and coherences are distributed over the sites 1, 2, 3, and 4, with sites 6, 7, and 8 being only weakly populated. This shows that the coherence length below 1 does not imply the lack of coherences but the possible existence of localized coherences, i.e. a small set of sites may have coherences between them. 


In Fig.~\ref{fig:corr_leng}B, when $\lambda \approx g$, the coherence length decreases for low temperatures in late stages of the dynamics. Inspecting  Fig.~\ref{fig:dm}E, we associate this decrease of the coherence length with localization of the exciton on the site 3, while retaining coherence with other sites. In the early stages of the dynamics, the exciton is originally at the site 1 and hops through other sites until it reaches site 3. An early increase in the coherence length can be associated with these initial steps of the dynamics. This is followed by  decreasing coherence length which is linked to the localization in other sites and finally at the site 3. At high temperatures, the coherence length shows a similar behavior to the case where the reorganization energy is high (Fig.~\ref{fig:corr_leng}C), as shown in  Fig.~\ref{fig:dm}B, whose reduced density matrix is similar to  Fig.~\ref{fig:dm}C. The difference amounts to very small correlations between sites, which explains 
different values of the coherence length: $L_{\rho}>1$ for Fig.~\ref{fig:dm}B  and $L_{\rho}<1$ for Fig.~\ref{fig:dm}C.  


For $\lambda = 40$ cm$^{-1}$ and low temperatures the exciton tends to  hop coherently between the sites until it localizes. In Fig.~\ref{fig:dm}D, the exciton is mainly localized over two sites, with only a few non-zero coherences. Figs.~\ref{fig:dm}D and E differ in the number of weak coherences that appear when  $\lambda < g$. Additionally, Fig.~\ref{fig:dm}A shows the behavior similar to Fig.~\ref{fig:dm}B in the high-temperature limit. 
Note that the coherence length is not bounded to 
2. 
Furthermore, in Fig.~\ref{fig:corr_low}\ref{apendix:lowlamda},  the coherence length 
 is shown to surpass the value of 2. 
Overall, we found that the coherence length does not exceed 4.


Moix \textit{et al.}\cite{moix2012equilibrium}  showed that the equilibrium coherence length has a maximum as a function of temperature for the 
FMO complex for a fixed reorganization energy and a cut-off frequency. 
Addressing this result would require, for some parameters, HEOM
simulations beyond our maximum run time of 2 ps, to guarantee that the equilibrium is reached. 
We, however, suspect that for $\lambda =40-160$ cm$^{-1}$ and $T\sim300$ K, the coherence length will 
decay, similar to what we observed for the low-temperature cases in the later stages of the dynamics. 

\subsection{ Efficiency }
\label{section:effi}



\begin{figure*}
\includegraphics[width=\textwidth]{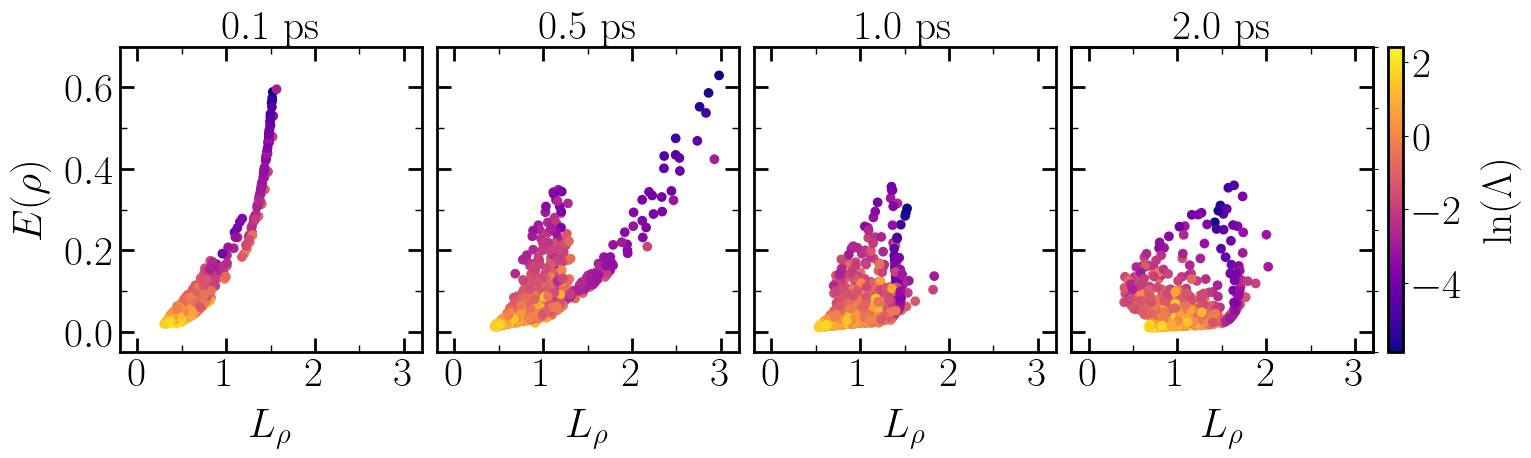}
\caption{\label{fig:corr_ent_leng} Global entanglement as a function of the coherence length for the entire data set of the reduced density matrices at different times of the dynamics.  Color of each point represents the efficiency parameter $\ln(\Lambda)$ (see main text for details).  }
\end{figure*}

Although the inter-exciton coherence are irrelevant for photosynthetic light-harvesting systems at physiological conditions, in this section we attempt to examine possible correlations between entanglement, coherence length, and efficiency.  To calculate the efficiency of energy transfer in the FMO complex, the irreversible energy flow from site 3 to the reaction center must be modeled through e.g., a Lindblad trapping operator, which is not included in our HEOM calculations. Although we cannot calculate the efficiency of the excitation energy transfer explicitly, we, again, use $\Lambda = \frac{\lambda}{\beta \gamma g }$ as a qualitative efficiency parameter.\cite{mohseni2014energy}  In Fig.~\ref{fig:corr_ent_leng}, we show the entanglement and the coherence length for all the parameter sets available in our database. Each point represents a set of parameters $(\gamma, \lambda, \beta)$, and different times 0.1, 0.5, 1.0, and 2.0 ps are shown in the corresponding panels. The color of each point represents the value of $\ln \Lambda$. 
At the early times of the dynamics, a correlation between coherence length and the entanglement is shown in Fig.~\ref{fig:corr_ent_leng}A. In this case, larger coherence length correlates with 
stronger entanglement. However, this behavior is lost, especially for $\ln \Lambda > 1$, where these points remain in the same region, and its evolution is plateaued.

In the region where the excitation energy transport is efficient, i.e. $\ln \Lambda = 0$, points do not expand in time considerably, showing that the exciton, in this case, is only partially localized with $0.5<L_{\rho}< 1.5 $ and the entanglement is weak. 
This might suggest that entanglement between sites does not play a direct role in the efficiency of the excitation energy transport.\cite{zerah2021photosynthetic} Exciton should be localized in the target site to be transferred to 
the reaction center. 

Furthermore, we believe that the lack of entanglement between the sites for $\ln \Lambda = 0$  in the FMO complex could imply high correlations between whole complex and the environment, which may be the critical factor for the high efficiency.


\section{Conclusions}
\label{section:conclu}


The FMO complex, like other light-harvesting  complexes, is remarkable in its 
efficiency of transferring the energy through the complicated network of BChla sites and the surrounding protein scaffold. 
Exciton dynamics in the FMO complex is a result of the interplay between coherent interaction between the sites and dissipation imposed by the bath. Here we study the global entanglement dynamics in the FMO complex across different environmental parameters. Global entanglement was found to be driven by the coexistence of non-Markovian effects and decoherence quantified by $\beta \gamma$. The amount of entanglement is controlled by $\beta \gamma$ for a fixed reorganization energy, but the overall shape of the entanglement dynamics is dictated by the reorganization energy $\lambda$. As expected, stronger dissipation suppresses entanglement.


Coherence length is driven by temperature and reorganization energy and is robust to non-Markovian effects. Exciton delocalization at early dynamics is associated with the initial steps of the energy transfer in the strongly coupled sites 1 and 2, in the case when the reorganization energy is similar or lower compared to the average energy $g$.  Furthermore, it was shown that the coherence length below 1 can imply quantum delocalization on a subsystem of the FMO complex, where entanglement can exist. Namely, a coherence length of less than 1 does not imply classical localization and lack of quantum correlations.  A clear connection between entanglement and efficiency was not found in agreement with previous works. 


The exciton dynamics does not furnish the explicit information about the environmental dynamics. However, as the interaction between the system and bath becomes stronger, the entanglement between them will grow. A further method, such as the reaction coordinate mapping,\cite{strasberg2016nonequilibrium,iles2014environmental} will be needed to corroborate this and unravel the true connections between quantum correlations and efficient excitation energy transfer, taking into account the protein environment.

\begin{acknowledgments}
This work is supported by the U.S. Department of Energy,
Office of Science, Office of Basic Energy
Sciences, under Award Number DE-SC0024511.
A.A.K. also acknowledges the start-up funds provided by the College of Arts and Sciences
and the Department of Physics and Astronomy of the University of Delaware,
and the General University Research award. Calculations were performed with high-performance
computing resources provided by the University of Delaware.
\end{acknowledgments}

\section*{Data Availability Statement}

The data set of FMO dynamics used in this work was published previously~\cite{ullah2023qd3set}   and can be accessed at \href{https://doi.org/10.25452/figshare.plus.c.6389553}{\url{https://doi.org/10.25452/figshare.plus.c.6389553}}.

\appendix

\section{FMO Hamiltonian }
\label{apendix:hamil}

The FMO Hamiltonian used in this work was taken from Ref.\citenum{ke2016hierarchy}, 

\begin{equation}
\left(\begin{array}{cccccccc}
310 & -80.3 & 3.5 & -4.0 & 4.5 & -10.2 & -4.9 & 21.0 \\
-80.3 & 230 & 23.5 & 6.7 & 0.5 & 7.5 & 1.5 & 3.3 \\
3.5 & 23.5 & 0 & -49.8 & -1.5 & -6.5 & 1.2 & 0.7 \\
-4.0 & 6.7 & -49.8 & 180 & 63.4 & -13.3 & -42.2 & -1.2 \\
4.5 & 0.5 & -1.5 & 63.4 & 450 & 55.8 & 4.7 & 2.8 \\
-10.2 & 7.5 & -6.5 & -13.3 & 55.8 & 320 & 33.0 & -7.3 \\
-4.9 & 1.5 & 1.2 & -42.2 & 4.7 & 33.0 & 270 & -8.7 \\
21.0 & 3.3 & 0.7 & -1.2 & 2.8 & -7.3 & -8.7 & 505
\end{array}\right),
\label{eq:FMO_hamil}
\end{equation}
where all site energies and couplings are in $\mathrm{cm}^{-1}$
and the diagonal energy offset is $12195 \mathrm{~cm}^{-1}$.

\section{Entanglement to small reorganization energies}
\label{apendix:lowlamda}

\begin{figure}
\includegraphics[width=0.5\textwidth]{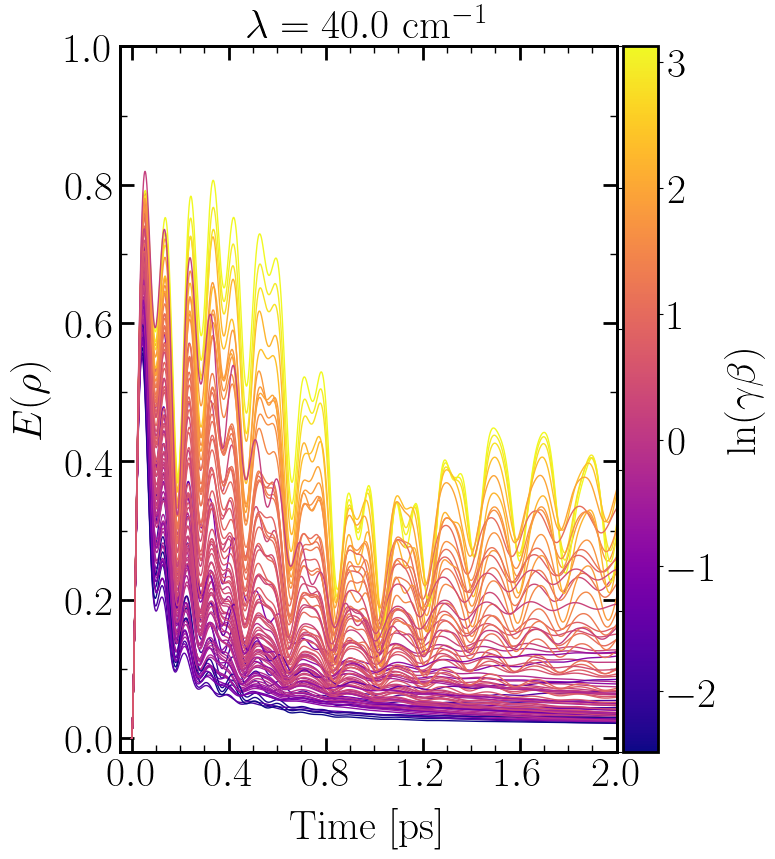}
\caption{\label{fig:low_globa}Global entanglement for $\lambda =10$ cm$^{-1}$. Color bar values correspond to the value of $\ln(\gamma \beta)$ same as  Figure \ref{fig:global-lambda}.}
\end{figure}

\begin{figure}
\includegraphics[width=0.5\textwidth]{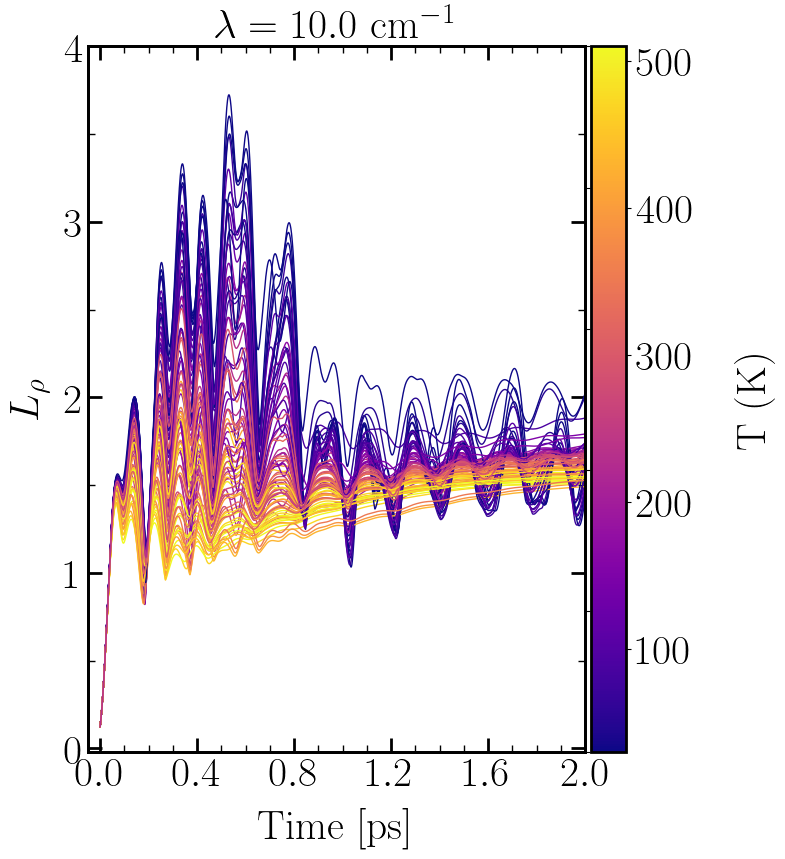}
\caption{\label{fig:corr_low} Coherence length for $\lambda =10$ cm$^{-1}$. Exciton shows high values of delocalization, and equilibrium coherence lengths are bigger than 1. }
\end{figure}

The reorganization energy controls the interaction of the FMO complex and its environment.  
When reorganization energy is small $\lambda \ll g $, the exciton dynamics is close to the unitary dynamics. Fig.~\ref{fig:low_globa} shows the global entanglement for $ \lambda=10$ cm$^{-1}$, where entanglement is persistent and oscillating. Additionally, Fig.~\ref{fig:corr_low}, shows that the exciton is mostly delocalized over
two sites $(1< L_{\rho} < 2) $  for all values of $\beta \gamma$. In the early stages of the dynamics, exciton could be delocalized over around four sites.

\section{Coherence length for a fixed cut-off frequency}
\label{apendix:cut-oof_ind}

\begin{figure}
\includegraphics[width=\textwidth]{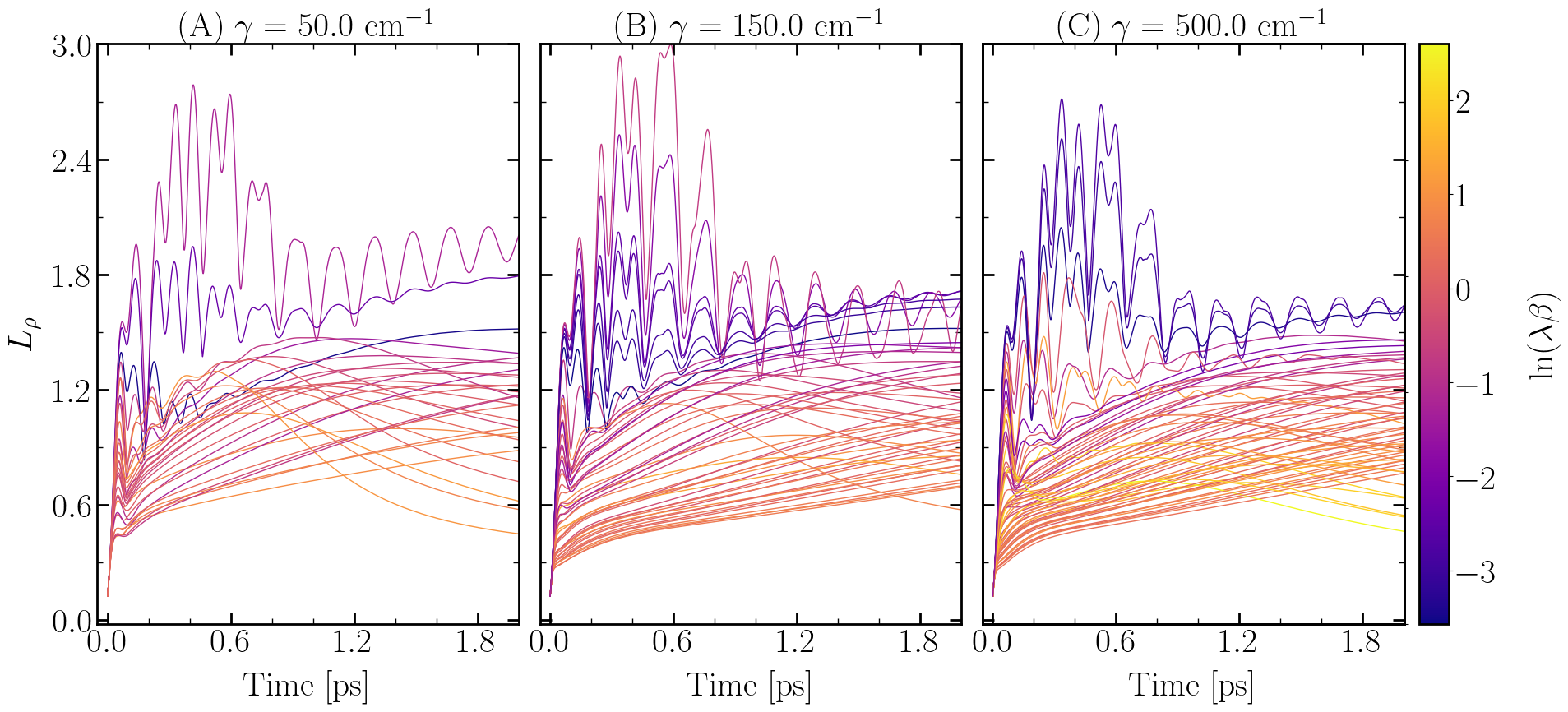}
\caption{\label{fig:gamma_fix} Coherence length for fixed values of cut-off frequency $\gamma=$ 50, 150, 500 cm$^{-1}$.}
\end{figure}

Fig.~\ref{fig:gamma_fix} shows that coherence length does not change appreciably when  the cut-off frequency is varied. 
Figs.~\ref{fig:gamma_fix}A-C show qualitatively similar behavior. 

\bibliography{aipsamp}

\end{document}